\documentclass[utf8]{FrontiersinHarvard} 
\usepackage{url,hyperref,lineno,microtype,subcaption}
\usepackage[onehalfspacing]{setspace}

\usepackage{bm} 

\def\keyFont{\fontsize{8}{11}\helveticabold }
\def\firstAuthorLast{Li et al.}
\def\Authors{Xinwei Li\,$^{1,2}$, Junichiro Kono\,$^{3,4,5}$ Qimiao Si\,$^{5}$
and Silke Paschen$^{6,*}$}
\def\Address{$^{1}$Department of Physics, California Institute of Technology, Pasadena, CA, 91125, USA \\
$^2$Institute for Quantum Information and Matter, California Institute of Technology, Pasadena, CA, 91125, USA \\
$^{3}$Department of Electrical and Computer Engineering, 6100 Main Street, Rice University, Houston, TX 77005, USA \\
$^{4}$Department of Materials Science and Nanoengineering, 6100 Main Street, Rice University, Houston, TX 77005, USA \\
$^{5}$Department of Physics and Astronomy, Center for Quantum Materials, 6100 Main Street, Rice University, Houston, TX 77005, USA \\
$^6$Institute of Solid State Physics, Technische Universit\"at Wien, Wiedner Hauptstra{\ss}e 8-10, 1040 Vienna, Austria}
\def\corrAuthor{Silke Paschen}
\def\corrEmail{paschen@ifp.tuwien.ac.at}

\begin{document}
\onecolumn
\firstpage{1}

\title[Planckian Optical Conductivity?]{Is the Optical Conductivity of Heavy Fermion Strange Metals Planckian?} 

\author[\firstAuthorLast ]{\Authors}
\address{} 
\correspondance{}

\extraAuth{}

\maketitle

\begin{abstract}

Strange metal behavior appears across a variety of condensed matter settings and
beyond, and achieving a universal understanding is an exciting prospect. The
beyond-Landau quantum criticality of Kondo destruction has had considerable
success in describing the behavior of strange metal heavy fermion compounds, and
there is some evidence that the associated partial localization-delocalization
nature can be generalized to diverse materials classes. Other potential
overarching principles at play are also being explored. An intriguing proposal
is that Planckian scattering, with a rate of $k_{\rm B}T/\hbar$, leads to the
linear temperature dependence of the (dc) electrical resistivity, which is a
hallmark of strange metal behavior. Here we extend a previously introduced
analysis scheme based on the Drude description of the dc resistivity to optical
conductivity data. When they are well described by a simple (ac) Drude model,
the scattering rate can be directly extracted. This avoids the need to determine
the ratio of charge carrier concentration to effective mass, which has
complicated previous analyses based on the dc resistivity. However, we point out
that strange metals typically exhibit strong deviations from Drude behavior, as
exemplified by the ``extreme'' strange metal YbRh$_2$Si$_2$. This calls for
alternative approaches, and we point to the power of strange metal dynamical
(energy-over-temperature) scaling analyses for the inelastic part of the optical
conductivity. If such scaling extends to the low-frequency limit, a strange
metal relaxation rate can be estimated, and may ultimately be used to test
whether strange metals relax in a Planckian manner.

\tiny \keyFont{ \section{Keywords:} strange metals, Planckian scattering,
 optical conductivity, Drude model, heavy fermion compounds, quantum
 criticality, non-Fermi liquid, YbRh$_2$Si$_2$}

\end{abstract}

\section{Introduction}\label{intro}

The standard theory of metals is Fermi liquid theory. It describes materials
across the correlation spectrum, from the simplest metals such as sodium or
aluminum to heavy fermion compounds \citep{Ste84.1} with mass renormalizations
of more than three orders of magnitude \citep{Kad86.1,Jac09.1}. However, under
certain circumstances, behavior at odds with this theory, dubbed non-Fermi
liquid behavior, is observed. From the perspective of Fermi liquid theory, this
regime arises when the quasiparticle weight factor $Z$, which decreases with
increasing correlation strength, vanishes. This is where Fermi liquid theory
breaks down and alternative descriptions are needed. In spite of tremendous
efforts and much progress, a full understanding of non-Fermi liquids is an
outstanding challenge.

One of the best understood settings in which non-Fermi liquid behavior arises
are continuous quantum phase transitions \citep{Sac99.1}. Heavy fermion
compounds have proven particularly suitable to study them because the competing
energy scales governing these materials make them readily tunable by external
control parameters \citep{Loe07.1,Kir20.1,Pas21.1,Pas21.2}. Through a large body
of investigations on many different compounds it became clear that two broad
categories can be distinguished. In the first one, the quantum critical
behavior, typically observed at the border of antiferromagnetic order in these
materials, is dominated by fluctuations of the order parameter. This leads to
strong scattering at hot spots on the Fermi surface that are associated with the
ordering wave vector, for which the resistivity in the clean case is
proportional to $T^2$ (where $T$ is the temperature) given that the current is
primarily carried by Landau quasiparticles that reside in the ``cold regions''
of the Fermi surface \citep{Hlu95.1}. In the presence of disorder, the
electrical resistivity is predicted to vary as $T^{\epsilon}$, with $1 \le
\epsilon \le 1.5$ depending on the amount of disorder \citep{Ros99.1}.
Predictions for such order parameter fluctuation or spin density wave (SDW)-type
quantum critical points (QCPs) have also been made  for other physical
properties \citep{Her76.1,Mil93.1,Mor95.1,Loe07.1,Zhu03.1}, and have been
verified in some cases \citep{Kue03.1}.

In the second category of materials, however, observations at odds with these
predictions have been made \citep{Pas21.1,Pas21.2}, which have provided support
for the beyond-Landau quantum criticality of Kondo destruction
\citep{Si01.1,Col01.1,Sen04.1}. These observations include
energy-over-temperature scaling of inelastic neutron scattering \citep{Sch00.1}
and optical conductivity data \citep{Pro20.1}, and an abrupt change of the Fermi
surface across the QCP as evidenced by de Haas--van Alphen \citep{Shi05.1} and
Hall effect measurements \citep{Pas04.1,Fri10.2,Cus12.1,Mar19.1}. The sudden
reconstruction from a small to a large Fermi surface characterizes a (partial,
$4f$ selective) localization-delocalization transition, and is emerging as a
potential universal organizing principle with supporting evidence in a variety
of other correlated materials classes \citep{Pas21.1}, including the cuprates
\citep{Bad16.1,Fan22.1}. The interacting nature of the fixed point describing a
Kondo-destruction QCP \citep{Si01.1,Col01.1,Sen04.1} means that $k_{\rm B}T$ is
the only energy scale, suggesting that the electrical resistivity would be
linearly proportional to $k_{\rm B}T$. Quite remarkably, all heavy fermion
compounds of this second category indeed exhibit a linear-in-temperature dc
electrical resistivity \citep{Tau22.1}. Thus, quantum criticality beyond order
parameter fluctuations appears to be tied to the ``strange metal''
linear-in-temperature resistivity in heavy fermion compounds.

The most striking case in this second category is YbRh$_2$Si$_2$, where the
electrical resistivity is linear in temperature over 3.5 orders of magnitude in
temperature, from above 10\,K down to the onset of unconventional
superconductivity near 3\,mK at the quantum critical field \citep{Ngu21.1}.
Recently, an attempt has been made \citep{Tau22.1} to answer the question
whether this extreme linear-in-temperature resistivity might be characterized by
``Planckian dissipation'' \citep{Zaa04.1}---scattering at a rate equal to the
inverse of the Planckian time scale (the shortest possible time scale in analogy
with the Planck time in quantum gravity). While the physics and implications of
the Planckian form of scattering remain unclear, its applicability to strange
metals is an intriguing question to ascertain empirically. As proposed in
\citep{Bru13.1} and more recently taken up by others
\citep{Leg19.1,Cao20.1,Ghi21.1,Gri21.1,Mou21.1}, this has been done by assuming
a Drude description of the (dc) resistivity. For YbRh$_2$Si$_2$, the scattering
rate was found to be much smaller, unless the charge carriers were assumed to
have unreasonably light masses. The same analysis was also carried out for other
strange metal heavy fermion compounds, with similar results \citep{Tau22.1}. As
such light carriers are absent in the heavy Fermi liquid ground state of these
materials, even at tuning parameter values far away from the quantum critical
value, it was concluded that dissipation was not Planckian \citep{Tau22.1}.
Nevertheless, a technique that could independently determine the effective mass
(more precisely, the ratio of charge carrier concentration to effective mass)
and the scattering rate is highly desirable. Here we show that the optical
conductivity can in principle accomplish this goal, albeit only if the data can
be described by the simple Drude form. As this is, however, typically not the
case in the strange metal regime, we discuss approaches based on dynamical
scaling as the way forward.

In this perspective paper, we start by introducing the Drude model of the
optical conductivity in the notation used here (section\,\ref{Drude}), and the
Planckian scattering analysis based on this model (section\,\ref{Planckian}).
Next, we use this description to first analyse the optical conductivity of
simple, non-interacting materials at high temperatures, where the scattering
from phonons leads to a linear-in-temperature dc resistivity
(sections\,\ref{Drude_simple}). Then we try to apply this Drude-based Planckian
scattering scheme to the optical conductivity of YbRh$_2$Si$_2$
(section\,\ref{Drude_YRS}). As anticipated, the strong deviation from Drude
behavior in the compound's strange metal regime limits this analysis to
temperatures and frequencies outside this regime, where Drude behavior is
recovered. Section\,\ref{nonDrude} is devoted to dynamical scaling analyses,
which are the appropriate tool to characterize strange metal optical
conductivities. In section\,\ref{RelaxationScaling}, we propose a scheme that
can determine a relevant strange metal relaxation rate---without resorting to a
Drude description---provided that data in the relevant temperature and frequency
regimes can be obtained. Before closing with a discussion and outlook
(section\,\ref{discussion}), we comment on the relation of the temperature and
frequency dependences of the optical conductivity in the Fermi liquid regime
(section\,\ref{Drude_UPA}).

\section{Drude formulation of the optical conductivity}\label{Drude}

In the Drude model, conduction electrons in solids are described as particles of
a classical gas executing diffusive motion with an average relaxation time
$\tau$. The equation of motion in the presence of a dc electric field leads to
the Drude form of the dc electrical conductivity
\begin{equation}
\sigma = \frac{n e^2 \tau}{m} \; , \label{eq:Drude_dc}
\end{equation}
where $n$ is the charge carrier concentration, $m$ their effective mass, and
$-e$ the electronic charge. In an ac field, a complex, frequency-dependent
optical conductivity
\begin{equation}
\sigma(\omega) = \frac{n e^2 \tau}{m} \frac{1}{1-i\omega \tau} = \frac{n e^2 \tau}{m} \frac{1 + i \omega \tau}{1 + \omega^2 \tau^2} \label{eq:Drude_ac}
\end{equation}
results, with the real part
\begin{equation}
{\rm Re}[\sigma(\omega)] = \sigma_1 = \frac{n e^2 \tau}{m} \frac{1}{1 + \omega^2 \tau^2} \label{eq:Drude_ac_Re}
\end{equation}
and the imaginary part
\begin{equation}
{\rm Im}[\sigma(\omega)] = \sigma_2 = \frac{n e^2 \tau}{m} \frac{\omega \tau}{1 + \omega^2 \tau^2} \; . \label{eq:Drude_ac_Im}
\end{equation}
We also introduce the plasma frequency
\begin{equation}
\omega_{\rm p} = \sqrt{\frac{n e^2}{m \epsilon_0}} \label{eq:plasma}
\end{equation}
with the permittivity of free space $\epsilon_0$.

Frequently, it is useful to discriminate between residual scattering due to
defects and intrinsic scattering. In the case of the dc resistivity, this is
what underlies the usually adopted forms
\begin{equation}
\rho = \rho_0 + A T^2 = \frac{1}{\sigma_{\rm res}} + \frac{1}{\sigma_{\rm in}} = \rho_{\rm res} + \rho_{\rm in} \label{eq:dc_FL}
\end{equation}
and
\begin{equation}
\rho = \rho_0' + A' T = \frac{1}{\sigma_{\rm res}} + \frac{1}{\sigma_{\rm in}} = \rho_{\rm res} + \rho_{\rm in} \label{eq:dc_SM}
\end{equation}
for Fermi liquids and strange metals, respectively. Here $\rho_0$ and $\rho_0'$
are the extrapolations to $T=0$ of a quadratic-in-temperature and a
linear-in-temperature dependence with slope $A$ and $A'$, respectively;
$\sigma_{\rm res}$ and $\sigma_{\rm in}$ (or $\rho_{\rm res}$ and $\rho_{\rm
in}$) are the zero-temperature residual (usually elastic) and the
temperature-dependent intrinsic (typically inelastic) contributions,
respectively. The underlying assumption is that the Matthiessen rule, which
states that for independent scattering channels the scattering rates and thus
the corresponding resistivities (or inverse conductivities) add, holds. Making
the same assumption for the optical conductivity leads to
\begin{equation}
\frac{1}{\sigma(\omega)} = \frac{1}{\sigma_{\rm res}(\omega)} + \frac{1}{\sigma_{\rm in}(\omega)} \; .\label{eq:ac_Matthiessen}
\end{equation}
Within the Drude approach, both $\sigma_{\rm res}$ and $\sigma_{\rm in}$ should
have the Drude form of equation \ref{eq:Drude_ac}, with $\tau = \tau_{\rm res}$
and $\tau = \tau_{\rm in}$, respectively.

In special cases, discussed further below, the situation may arise that the
residual scattering rate is much smaller than the intrinsic one, such that
\begin{equation}
\frac{1}{\tau} = \frac{1}{\tau_{\rm res}} + \frac{1}{\tau_{\rm in}} \approx \frac{1}{\tau_{\rm in}} \; . \label{eq:tau_intrinsic}
\end{equation}
One can then rewrite equation \ref{eq:Drude_ac} as
\begin{equation}
\sigma(\omega) = \frac{ne^2}{m}\frac{\tau}{(1-i\omega\tau)} = \frac{ne^2}{m}\frac{1}{(1/\tau-i\omega)} \approx \frac{ne^2}{m} \frac{1}{(1/\tau_\text{in}-i\omega)}=\sigma_\text{in}(\omega) \; . \label{eq:ac_intrinsic}
\end{equation}
 
\section{Drude analysis of the optical conductivity for Planckian scattering analysis}\label{Planckian}

Previous attempts to characterize strange metal behavior in terms of Planckian
scattering using dc conductivity data faced the problem that the scattering time
$\tau$ appears in a product with the ratio of charge carrier concentration to
mass, $n/m$; see equation \ref{eq:Drude_dc}. Thus, to estimate $\tau$ or, more
precisely, $\tau_{\rm in}$ and compare it with the Planckian time
\begin{equation}
\tau_{\rm P} = \frac{\hbar}{k_{\rm B} T} \; , \label{eq:tau_Pl}
\end{equation}
as typically done by quantifying the coefficient
\begin{equation}
\alpha =  \frac{\tau_{\rm P}}{\tau_{\rm in}} =  \frac{\hbar}{k_{\rm B} T}\frac{1}{\tau_{\rm in}} \equiv \alpha_{\tau}\; , \label{eq:alpha}
\end{equation}
required the knowledge of $n/m$, as seen explicitly by inserting equations
\ref{eq:Drude_dc} and \ref{eq:dc_SM} to obtain
\begin{equation}
\alpha =  \frac{n}{m}\frac{e^2\hbar}{k_{\rm B}}A' \equiv \alpha_{n/m} \; . \label{eq:alpha1}
\end{equation}
Because extracting $n$ and $m$ from different physical quantities gave different
results, this led to conflicting conclusions, as discussed in \citep{Tau22.1}.

Here we propose an analysis of the optical conductivity that can, in principle,
avoid this problem. As can be seen from equation \ref{eq:Drude_ac}, in addition
to appearing in a product with $n/m$ (in the prefactor), $\tau$ enters as the
only parameter in the second factor of $\sigma(\omega)$. Thus, fitting the Drude
form to the (intrinsic) optical conductivity data allows one to extract
$\tau_{\rm in}$ and $n/m$ independently. One can therefore determine $\alpha$
directly via equation \ref{eq:alpha} or, in combination with dc data in the
linear-in-temperature regime (which gives $A'$), using equation \ref{eq:alpha1}.
We once again emphasize that this analysis scheme requires the data to have the
Drude form. We will come back to this important point in
section\,\ref{Drude_YRS}.

\section{Drude-like optical conductivity of lead, aluminum, and silicon}\label{Drude_simple}

As discussed in section\,\ref{intro}, strange metal behavior refers to the
unusual properties exhibited by a number of strongly correlated electron systems
at low temperatures, most notably a linear-in-temperature electrical resistivity
where a normal metal would exhibit Fermi liquid behavior. However, this
temperature dependence can also arise from entirely different physics, that of
electrons scattered by classical (macroscopically populated) phonons at
sufficiently high temperatures. This case was included in a previous Planckian
dissipation analysis \citep{Bru13.1}, and we revisit it here with two metals,
lead and aluminum, from the perspective of the optical conductivity. Combining
the $A'$ coefficients (0.071\,$\mu\Omega$cm/K and 0.01\,$\mu\Omega$cm/K,
respectively) of their linear-in-temperature dc resistivities at high
temperatures with $n/m$ values estimated from quantum oscillation data, $\alpha$
values of 2.7 and 1.1 were obtained \citep{Bru13.1}.

\begin{figure}[t!]
\begin{center}
\includegraphics[width=18cm]{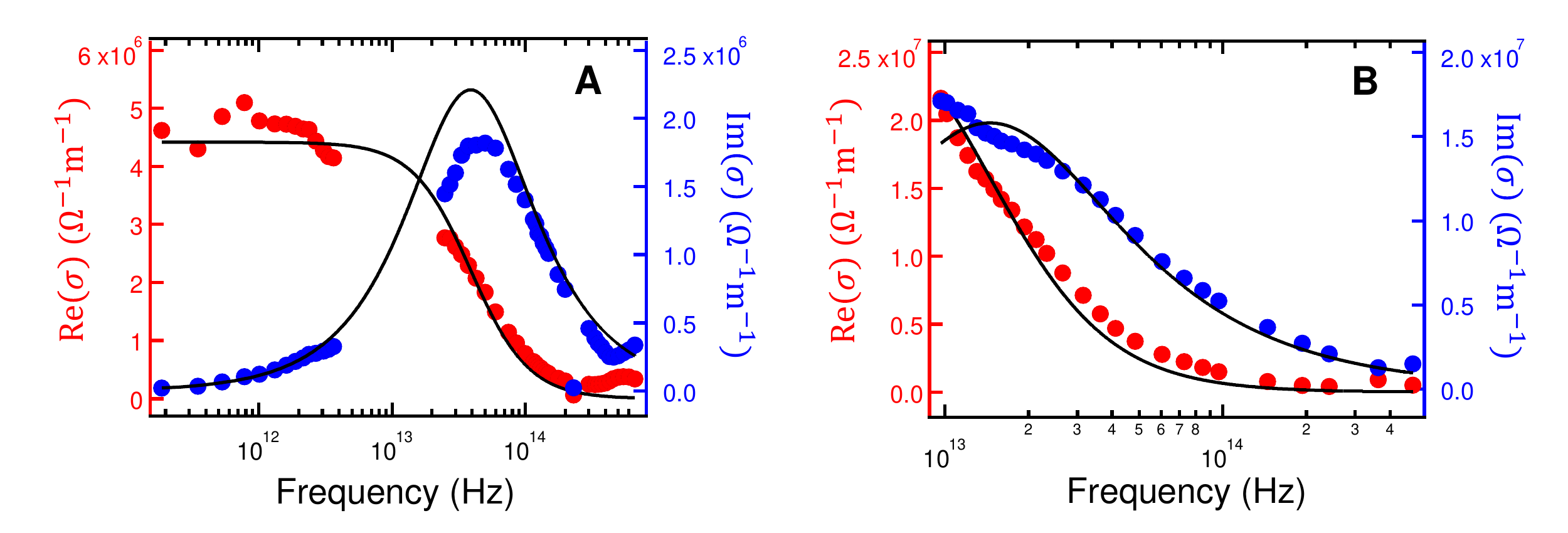}
\end{center}
\vspace{-0.5cm}

\caption{Drude fits to the optical conductivity of \textbf{(A)} lead (Pb) and
\textbf{(B)} aluminum (Al) at room temperature. The real and imaginary parts of
the conductivity are represented as red and blue markers, respectively. The
Drude fits are shown in black solid lines. See text for the results. The raw
data were obtained from
\citep{Ord83.1,Shi80.1,Bra72.1,Gol68.1}.}\label{fig:PbAl}
\end{figure}

Here we use tabulated optical conductivity data at 300\,K
\citep{Ord83.1,Shi80.1,Bra72.1,Gol68.1} and fit the real and imaginary parts
using equations \ref{eq:Drude_ac_Re} and \ref{eq:Drude_ac_Im} to obtain a unique
parameter set of $n/m$ and $\tau$ (figure \ref{fig:PbAl}); the process minimizes
the joint standard deviation, which is calculated by adding up the fitting
residuals associated with both the real and imaginary parts. Because lead and
aluminum are good metals, we can assume equations \ref{eq:tau_intrinsic} and
\ref{eq:ac_intrinsic} to be fulfilled and thus used the total conductivity
without residual resistivity corrections.

For lead, we obtain $\tau=4.08\cdot 10^{-15}$\,s/rad and $n/m=4.23\cdot
10^{58}$\,m$^{-3}$kg$^{-1}$, which gives $\alpha_{\tau}=6.3$ and
$\alpha_{n/m}=5.9$ (with the above value for $A'$). Using the same method for
aluminum, we obtain $\tau=1.10\cdot 10^{-14}$\,s/rad and $n/m=1.13\cdot
10^{59}$\,m$^{-3}$kg$^{-1}$, which gives $\alpha_{\tau}=2.3$ and
$\alpha_{n/m}=2.2$. The good agreement between $\alpha_{\tau}$ and
$\alpha_{n/m}$ indicates internal consistency. For both metals, these $\alpha$
values are about two times larger than the values obtained in \citep{Bru13.1},
pointing to some inaccuracy in the estimation of $n/m$ from quantum oscillation
experiments even for simple metals.

\begin{figure}[b!]
\begin{center}
\includegraphics[width=12cm]{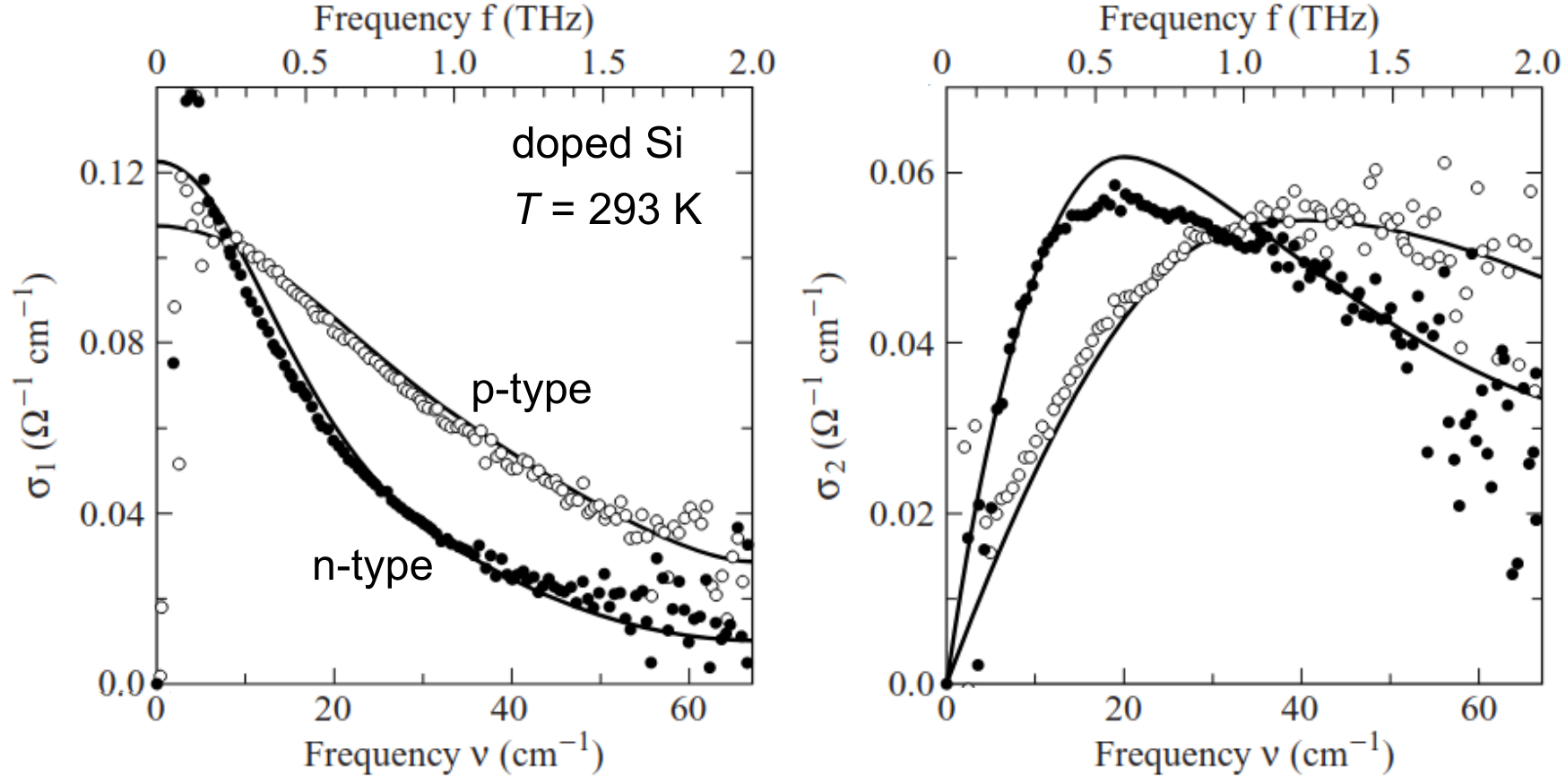}
\end{center}
\vspace{-6.3cm}

\hspace{3cm}{\large\bf{\fontfamily{phv}\selectfont A}}\hspace{5.7cm}{\large\bf{\fontfamily{phv}\selectfont B}}
\vspace{5.8cm}

\caption{Frequency dependence of the \textbf{(A)} real and \textbf{(B)}
imaginary parts of the optical conductivity of silicon, weakly doped by holes
($1.1 \cdot 10^{15}$\,cm$^{-3}$, open circles) or electrons ($4.2 \cdot
10^{14}$\,cm$^{-3}$, solid circles). The experiments were performed at $T =
293$\,K in the time domain; the full lines correspond to Drude fits (see text).
Figure adapted from \citep{Dre02.2}, with original data and Drude fits from
\citep{Ext90.1}.}\label{fig:Si}
\end{figure}

For comparison, we also include published optical conductivity data of
moderately doped silicon \citep{Ext90.1} (figure \ref{fig:Si}). At room
temperature, where the scattering in such samples is known to be dominated by
electron-phonon scattering, the scattering times $\tau$ obtained from the Drude
fits in \citep{Ext90.1} are $1.33\cdot 10^{-13}$\,s/rad and  $2.65\cdot
10^{-13}$\,s/rad for p- and n-type samples of the carrier concentrations
$1.1\cdot 10^{15}\,{\rm cm}^{-3}$ and $0.42\cdot 10^{15}\,{\rm cm}^{-3}$ and the
effective masses $0.37 m_0$ and $0.26 m_0$, respectively \citep{Ext90.1}, where
$m_0$ is the free electron mass. This results in $\alpha_{\tau}=0.19$ and
$0.096$, respectively, at least by a factor of five below the Planckian limit.

Our results for both simple metals and simple semiconductors thus indicate that
the scattering of electrons by classical phonons can deviate substantially from
Planckian scattering (where $\alpha \approx 1$).

\section{Drude contribution to the optical conductivity of
YbRh$_2$Si$_2$}\label{Drude_YRS}

Next we address the optical conductivity of the heavy fermion compound
YbRh$_2$Si$_2$. Strong correlations are known to push the Drude response of
metals to low frequencies, frequently outside the range of standard
spectrometers, making experiments on these materials challenging. When optical
reflectivity measurements are used, the Kramers--Kronig transformation is needed
to extract the optical conductivity. This may induce considerable uncertainty at
low frequencies. These problems were overcome in a recent study on
YbRh$_2$Si$_2$ thin films grown by molecular beam epitaxy (MBE) and measured by
terahertz (THz) time-domain transmission spectroscopy \citep{Pro20.1}. The data
were shown to exhibit dynamical scaling in the strange metal regime, which we
will come back to in section\,\ref{nonDrude}. Here, we first examine whether a
Drude description is possible in any frequency and temperature range such that
the above Planckian dissipation analysis can be performed.

\begin{figure}[t!]
\begin{center}
\includegraphics[width=18cm]{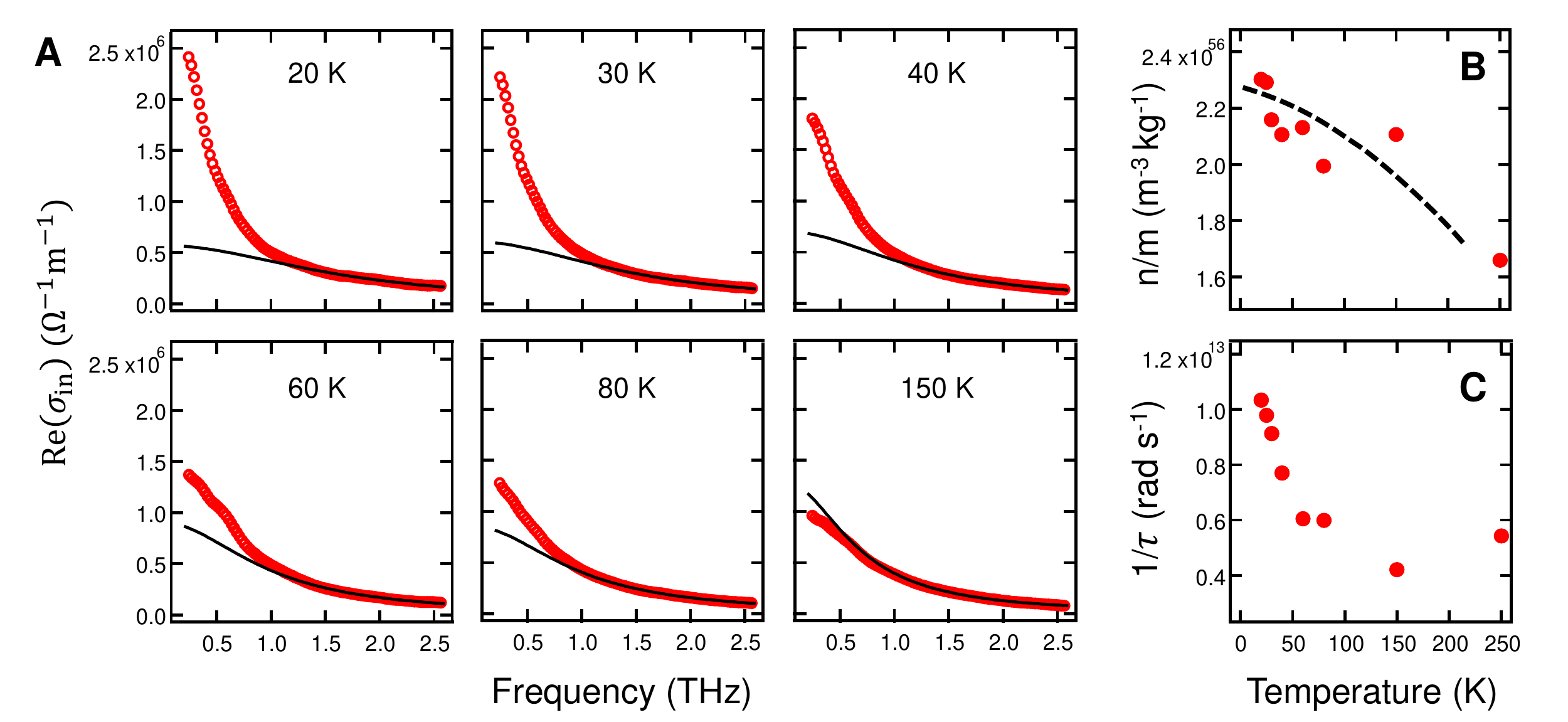}
\end{center}
\caption{\textbf{(A)} Drude fits to the real part of the intrinsic optical
conductivity of YbRh$_2$Si$_2$ at various temperatures. The subtraction of the 
residual resistivity is described in \citep{Pro20.1}. The fits (black solid
lines) are anchored to the high-frequency range of the experimental data (red
circles). Strong deviations from Drude behavior are seen at low frequencies, in
particular for the low-temperature isotherms. Temperature dependence of the
ratio $n/m$ \textbf{(B)} and the scattering rate $1/\tau$ \textbf{(C)} extracted
from the Drude fits. The black dashed line is a guide to the eyes for
extrapolating the $n/m$ value to $T=0$.}\label{fig:YRS_Drude}
\end{figure}

As shown in Fig.\,S3 of the Supplementary Material of \citep{Pro20.1}, only the
highest-temperature ${\rm Re}[\sigma(\omega)]$ data of YbRh$_2$Si$_2$ are well
described by a simple Drude model across the entire available frequency range
(0.23 to 2.6 THz). At lower temperatures, pronounced deviations appear at low
frequencies. In figure\,\ref{fig:YRS_Drude}A, we replot the real part of the
frequency dependent intrinsic optical conductivity from \citep{Pro20.1} at
different fixed temperatures between 20 and 150\,K. The black lines are Drude
fits to data at frequencies above 1.5\,THz. Deviations from Drude behavior at
lower  frequencies become stronger with decreasing temperature. Below 20\,K,
they are so strong that such fits lose any significance (and are therefore
discarded). Thus, the Drude description works only {\em outside} the material's
strange metal regime, in terms of both temperature and frequency. In the case of
the dc resistivity, a non-thermal tuning parameter was used to tune the material
into its Fermi liquid (and Drude) regime \citep{Tau22.1}; to do the same with
optical conductivity would require data at lower temperatures and
frequencies---an interesting study for future work. The  parameters $n/m$ and
$1/\tau$ extracted from our high-temperature and high-frequency Drude fits are
shown as a function of temperature in figure\,\ref{fig:YRS_Drude}B and C,
respectively. The parameter $n/m$ appears to saturate to $2.3 \cdot
10^{56}\,{\rm m}^{-3}{\rm kg}^{-1}$ at the lowest temperatures, which
corresponds to a plasma frequency of $0.54\,{\rm eV}$.  This is much larger than
the expectation of about $45\,{\rm meV}$ for the background (not quantum
critical) heavy Fermi liquid state, as estimated from $\sqrt{D k_{\rm B}T_{\rm
K}}$ \citep{Mil87.1}, where $D \sim 1\,{\rm eV}$ is the (bare) conduction
electron bandwidth and $T_{\rm K} =25\,{\rm K}$ is the single ion Kondo
temperature of YbRh$_2$Si$_2$ \citep{Tro00.2note}. Indeed, in the Fermi liquid
heavy fermion metal UPd$_2$Al$_3$, a plasma frequency of 41\,meV could be
extracted from very low-frequency optical (microwave) conductivity data, which
confirms this expectation. Using equation\,\ref{eq:alpha1} with $A' =
1.42\,\mu\Omega{\rm cm}$, the slope of the linear-in-temperature dc resistivity
of the MBE film used for the THz spectroscopy experiments, yields $\alpha_{n/m}
= 0.64$. On the other hand, using equation\,\ref{eq:alpha} and the result for
the scattering rate at 20\,K, $1/\tau = 1.03\cdot 10^{13}\,{\rm rad/s}$, yields
$\alpha_{\tau} = 3.93$, which is more than a factor of 6 larger than
$\alpha_{n/m}$. Compared to the cases of Pb and Al, the discrepancy between
$\alpha_{n/m}$ and $\alpha_{\tau}$ is significant. Furthermore, the resulting
temperature dependence of the scattering rate (figure\,\ref{fig:YRS_Drude}C) is
at odds with expectations for a strange metal. We conclude that a Drude analysis
of the optical conductivity fails to capture any pertinent aspect of the strange
metal behavior of YbRh$_2$Si$_2$.

\begin{figure}[t!]
\begin{center}
\includegraphics[width=12cm]{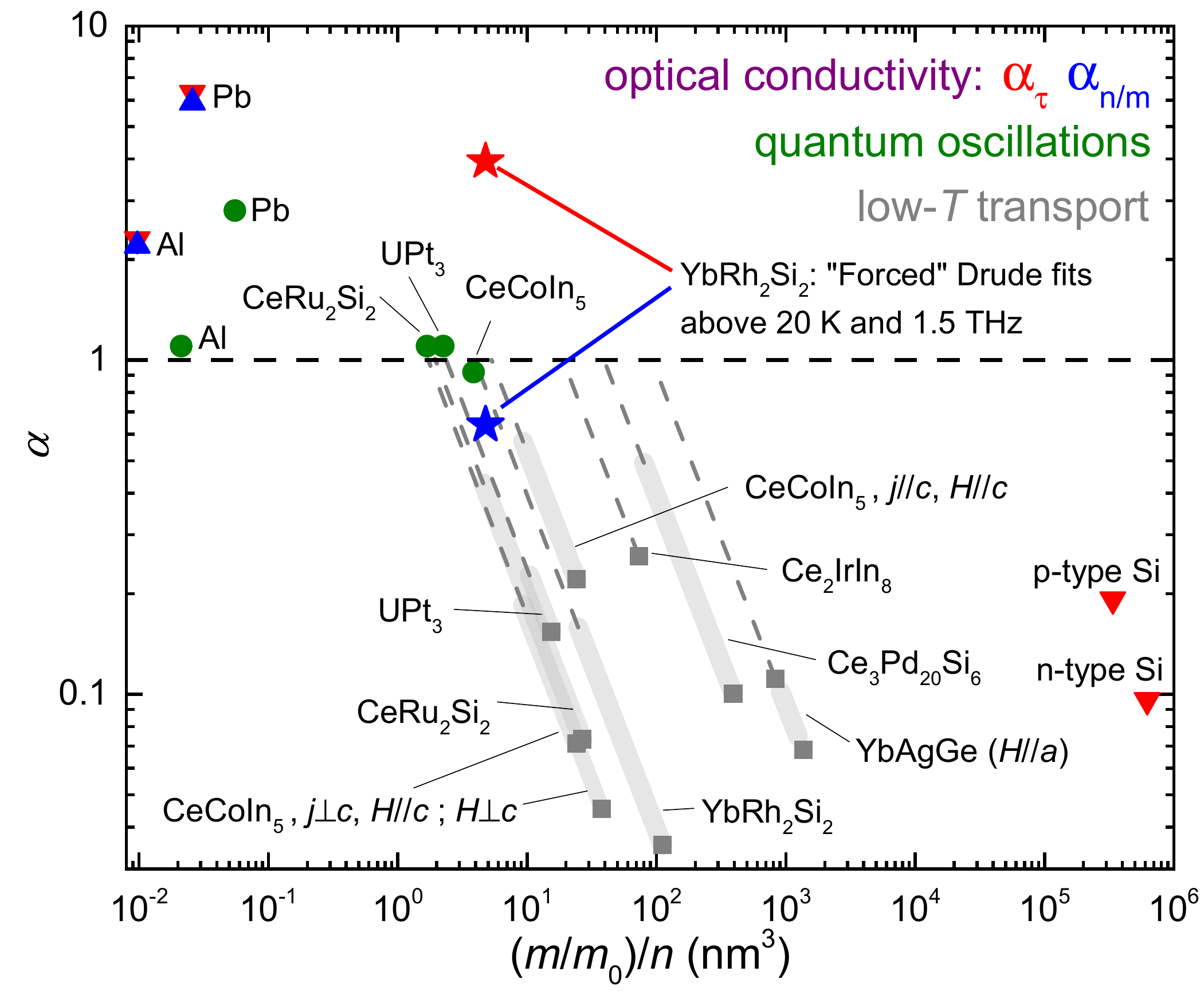}
\end{center}
\caption{Results of various Planckian analyses, in a double-logarithmic plot of
$\alpha$ vs $(m/m_0)/n$. The dark grey squares and the broad light grey lines
were obtained in \citep{Tau22.1} from low-temperature transport measurements.
The dashed grey lines are extrapolations to the dashed black $\alpha = 1$ line
of Planckian dissipation. Note that the grey lines---$\alpha$ vs $(m/m_0)/n$
curves obtained from equation\;\ref{eq:alpha1}---are straight lines of slope
$-1$ on this double-logarithmic plot, with offsets proportional to $A'$. The
green circles are $\alpha_{n/m}$ values obtained in \citep{Bru13.1} from quantum
oscillation data. The red and blue symbols are $\alpha_{\tau}$ and
$\alpha_{n/m}$, respectively, determined in the present work from Drude fits to
published optical conductivity data (see text). The fits for Pb, Al, and Si are
of good quality in the temperature range where the  scattering rates are
linear-in-temperature due to electron-phonon scattering. For YbRh$_2$Si$_2$, by
contrast, Drude fits fail in the strange metal regime (below 10 to 15\,K in the
dc resistivity). The $\alpha$ values (stars) were therefore extracted from Drude
fits to data above 20\,K and 1.5\,THz, where the fit quality is reasonable. The
discrepancy between $\alpha_{\tau}$ and $\alpha_{n/m}$ is large in this case,
which casts doubt on their significance.}\label{fig:alpha}
\end{figure}

We summarize the results of our Drude-based Planckian optical conductivity
analyses in figure\,\ref{fig:alpha}, which is an expanded version of Fig.\,4 of
\citep{Tau22.1}. The dark grey squares and broad light grey lines were obtained
in \citep{Tau22.1} by combining the effective mass (determined from the $A$
coefficient of the low-temperature dc electrical resistivity using the
Kadowaki--Woods ratio) and the charge carrier concentration (from Hall effect
measurements) with the $A'$ coefficient via equation\,\ref{eq:alpha1}. For each
material, the square represents the data point closest to the QCP, where the
effective mass is largest, and the broad line represents the range of $A$ values
measured away from the QCP, with the top end corresponding to the largest tuning
parameter separation from the quantum critical value (where the effective mass
is smallest). As discussed in \citep{Tau22.1}, it is clear that the $\alpha$
values obtained within this Drude analysis scheme are much smaller than 1,
suggesting that the strange metal resistivity of quantum critical heavy fermion
compounds cannot be understood as heavy quasiparticles undergoing Planckian
dissipation. Also included are $\alpha_{n/m}$ data of selected materials
obtained in \citep{Bru13.1} using published quantum oscillation data (green
points). The effective masses of these results are much smaller than those
determined through the resistivity, suggesting that quantum oscillation
experiments fail to detect the heaviest masses. The $\alpha$ values from
\citep{Bru13.1} thus do not represent the heavy (or heaviest) quasiparticles of
the respective compounds, but only lighter ones. This is supported by the fact
that for UPt$_3$, the large effective mass renormalization found in transport
and thermodynamic measurements was early on confirmed by optical conductivity
measurements to be above 160 \citep{Deg97.1}, whereas the value found in
\citep{Bru13.1} is less than 50.

The results of the present analyses are shown as red and blue symbols. The
ambiguity of which effective mass and which charge carrier concentration to use
is removed in the optical conductivity analysis because both $\tau$ and $n/m$
are obtained from the Drude fits. For the simple metals (Pb and Al) and
semiconductors (p- and n-type Si) at room temperature, where the resistivities
are governed by linear-in-temperature electron-phonon scattering rates,
$\alpha_{n/m}$ and $\alpha_{\tau}$ agree well (when both are available),
confirming that the analysis is robust. At least some of the $\alpha$ values are
sizably larger than 1, showing that the Planckian bound is not strictly obeyed
in the phonon scattering case. For YbRh$_2$Si$_2$, there is a large discrepancy
between $\alpha_{n/m}$ and $\alpha_{\tau}$, indicating that the analysis is not
meaningful. As discussed above, the Drude fits describe the optical conductivity
only at high temperatures and frequencies, but not in the strange metal regime.
The large $\alpha_{\tau}$ value is likely dominated by phonons. The fact that
$\alpha_{n/m}$ is much larger than the largest $\alpha$ value from the dc
resistivity analysis (top end of light grey line) indicates that it describes
only weakly renormalized quasiparticles with Drude behavior. 

An interesting observation is that the extrapolated zero-temperature value of
$n/m$ obtained from our Drude analysis ($2.3 \cdot 10^{56}\,{\rm m}^{-3}{\rm
kg}^{-1}$; see figure\,\ref{fig:YRS_Drude}B) is very similar to the value of
$2.2 \cdot 10^{56}\,{\rm m}^{-3}{\rm kg}^{-1}$ determined in \citep{Pro20.1} for
the residual resistivity. There, it was calculated from $n = 2.6 \cdot
10^{28}\,{\rm m}^{-3}$ as determined in \citep{Pas04.1} from low-temperature
Hall effect measurements and $m/m_0 \approx 130$ as determined from a deviation
($\chi^2$) minimization procedure performed within the dynamical scaling
analysis of the data [see Fig.\,S5 of the Supplementary Material of
\citep{Pro20.1}]. This may indicate that the quasiparticles described by our
Drude fits to ${\rm Re}[\sigma_{\rm in}(\omega)]$ are the same as the ones that
undergo residual (likely elastic) scattering. Their $n/m$ ratios suggest only
moderate mass renormalization ($130$ if the single-band interpretation of the
Hall coefficient is trusted). The much larger mass renormalizations as extracted
from the electronic specific heat coefficient $\gamma$ or the linear-in-$T^2$
resistivity coefficient $A$ via the Kadowaki--Woods ratio, as done in
\citep{Tau22.1}, thus appear to be entirely dynamically generated, and to be
part of the non-Drude regime in the optical conductivity even though extracted
from Fermi liquid relations of the specific heat and dc resistivity.

\section{Dynamical scaling of the intrinsic optical conductivity of strange metals}\label{nonDrude}

\begin{figure}[b!]
\begin{center}
\includegraphics[height=6cm]{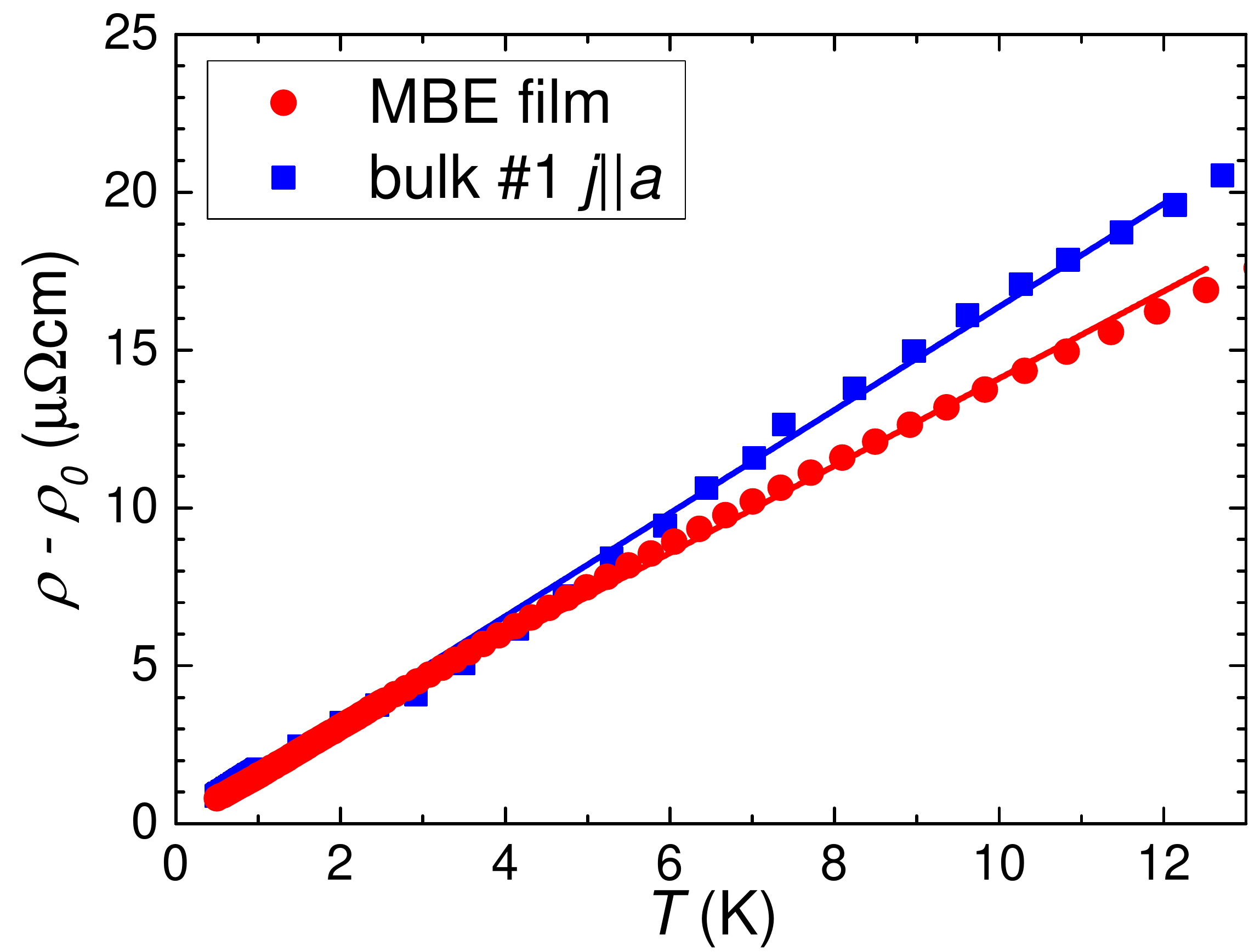}\hspace{0.6cm}\includegraphics[height=5.8cm]{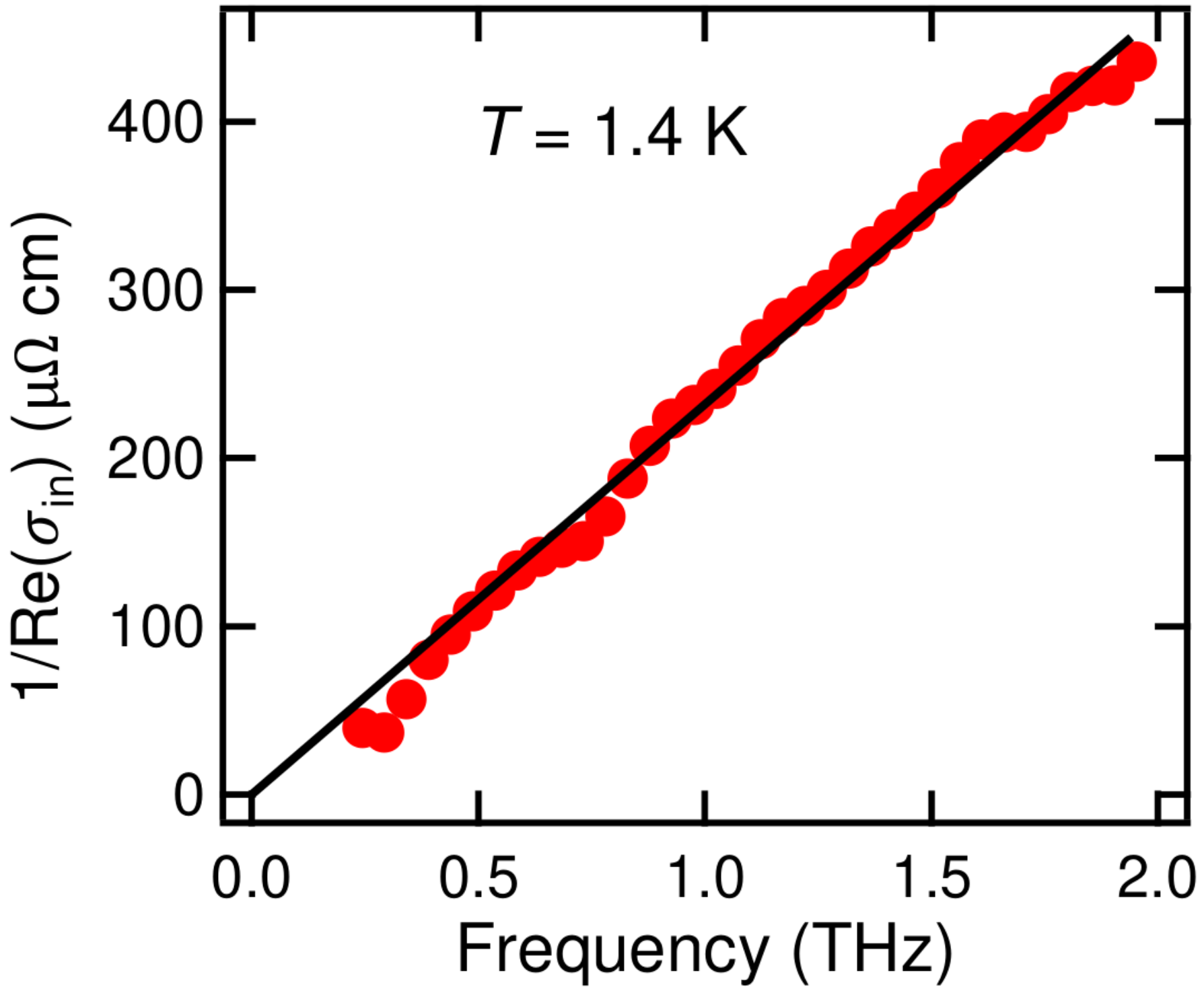}
\end{center}
\vspace{-6.2cm}

\hspace{1.2cm}{\large\bf{\fontfamily{phv}\selectfont A}}\hspace{8.0cm}{\large\bf{\fontfamily{phv}\selectfont B}}
\vspace{5.7cm}

\caption{Dc and THz resistivity of YbRh$_2$Si$_2$. \textbf{(A)} The dc
resistivity is linear in temperature. The slopes of the curve for an MBE film
and a bulk single crystal are similar. \textbf{(B)} The inverse of the real part
of the low-temperature intrinsic optical conductivity, 1/Re$(\sigma_{\rm in})$,
of the MBE film is approximately linear in frequency. Figures from
\citep{Pro20.1}.}\label{fig:YRS_1}
\end{figure}

As seen from figure\,\ref{fig:YRS_Drude} and also demonstrated in
\citep{Pro20.1}, ${\rm Re}[\sigma_{\rm in}(\omega)]$ of YbRh$_2$Si$_2$ shows
pronounced deviations from Drude behavior at low temperatures and frequencies.
These deviations appear to be related to the strange metal behavior of
YbRh$_2$Si$_2$, evidenced early on by the low-temperature upturn of the
electronic specific heat coefficient setting in somewhat below 30\,K and the
linear-in-temperature dc (intrinsic) resistivity $\rho_{\rm in}$ below 10 to
15\,K \citep{Tro00.2note}. Indeed, at the lowest temperature of 1.4\,K reached
in the THz spectroscopy experiments \citep{Pro20.1}, which corresponds to a
frequency $k_{\rm B}T/h = 0.03\,{\rm THz}$ well below the lowest accessed
frequency of 0.23\,THz, $1/{\rm Re}[\sigma_{\rm in}(\omega)]$ is linear in
frequency just as $\rho_{\rm in}$ is linear in temperature
(figure\,\ref{fig:YRS_1}).

One can ask whether the coefficients of the two linear dependences are related,
perhaps in a similar way to what is expected for Fermi liquids (see
section\,\ref{Drude_UPA}), namely
\begin{equation}
\frac{1}{\tau} = \frac{1}{\tau_{\rm res}} + \sqrt{a'^2 (k_{\rm B}T)^2 + b'^2 (\hbar \omega)^2}\; . \label{eq:NFL}
\end{equation} 
From the data in figure\,\ref{fig:YRS_1} we determine the ratio of $a'/b'$ to be
about 0.3. In analogy with the ratio $a/b = 4 \pi^2$ predicted for the Fermi
liquid case (section\,\ref{Drude_UPA}) one might have expected $a'/b' =
\sqrt{a/b} = 2 \pi$, which is, however, sizably larger than 0.3. Interestingly,
a similar discrepancy appears to occur even for Fermi liquids (see
section\,\ref{Drude_UPA}), calling for future work to reexamine the theoretical
expectation.

\begin{figure}[b!]
\begin{center}
\includegraphics[height=6.85cm]{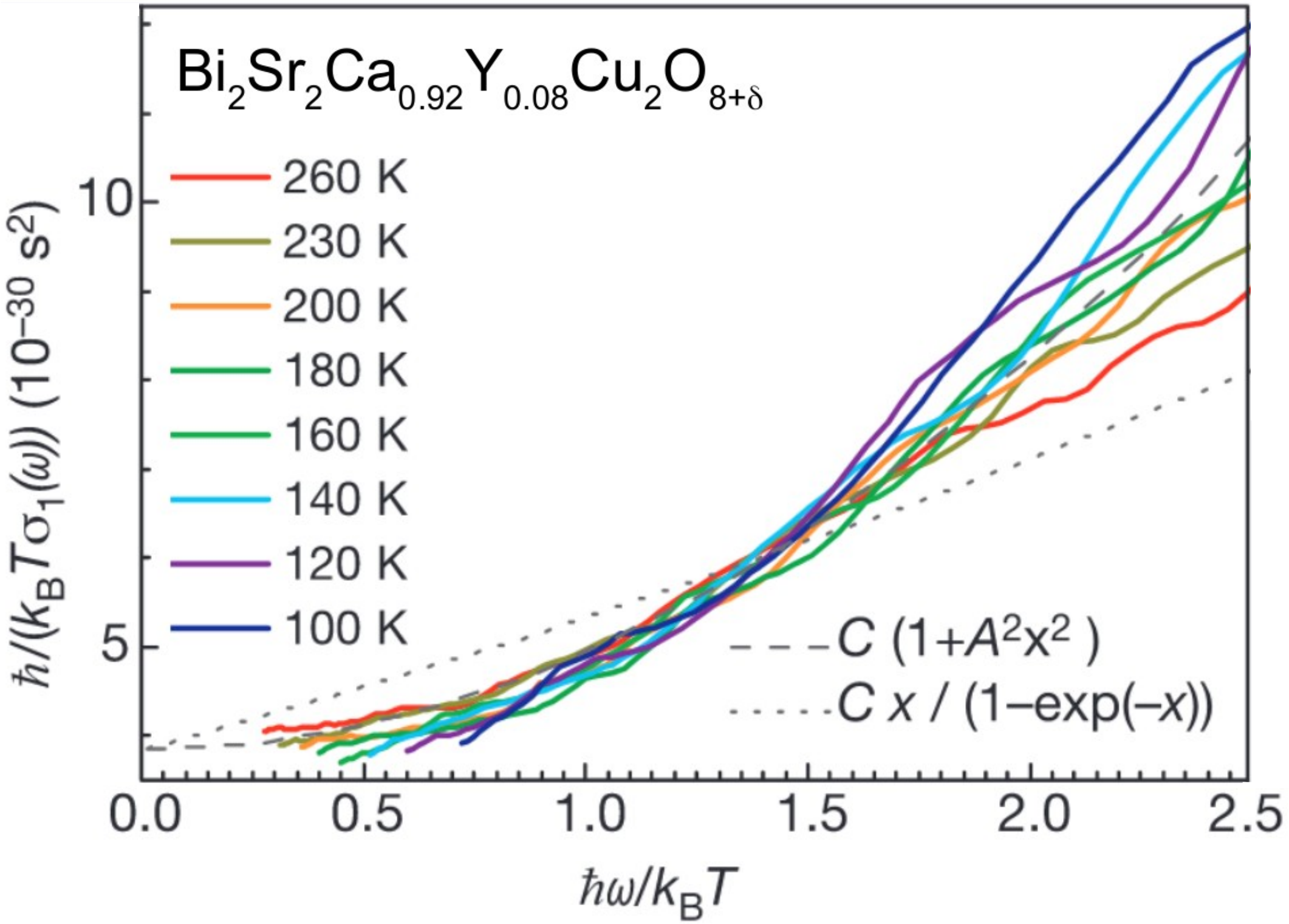}\hspace{0.2cm}\includegraphics[height=6.85cm]{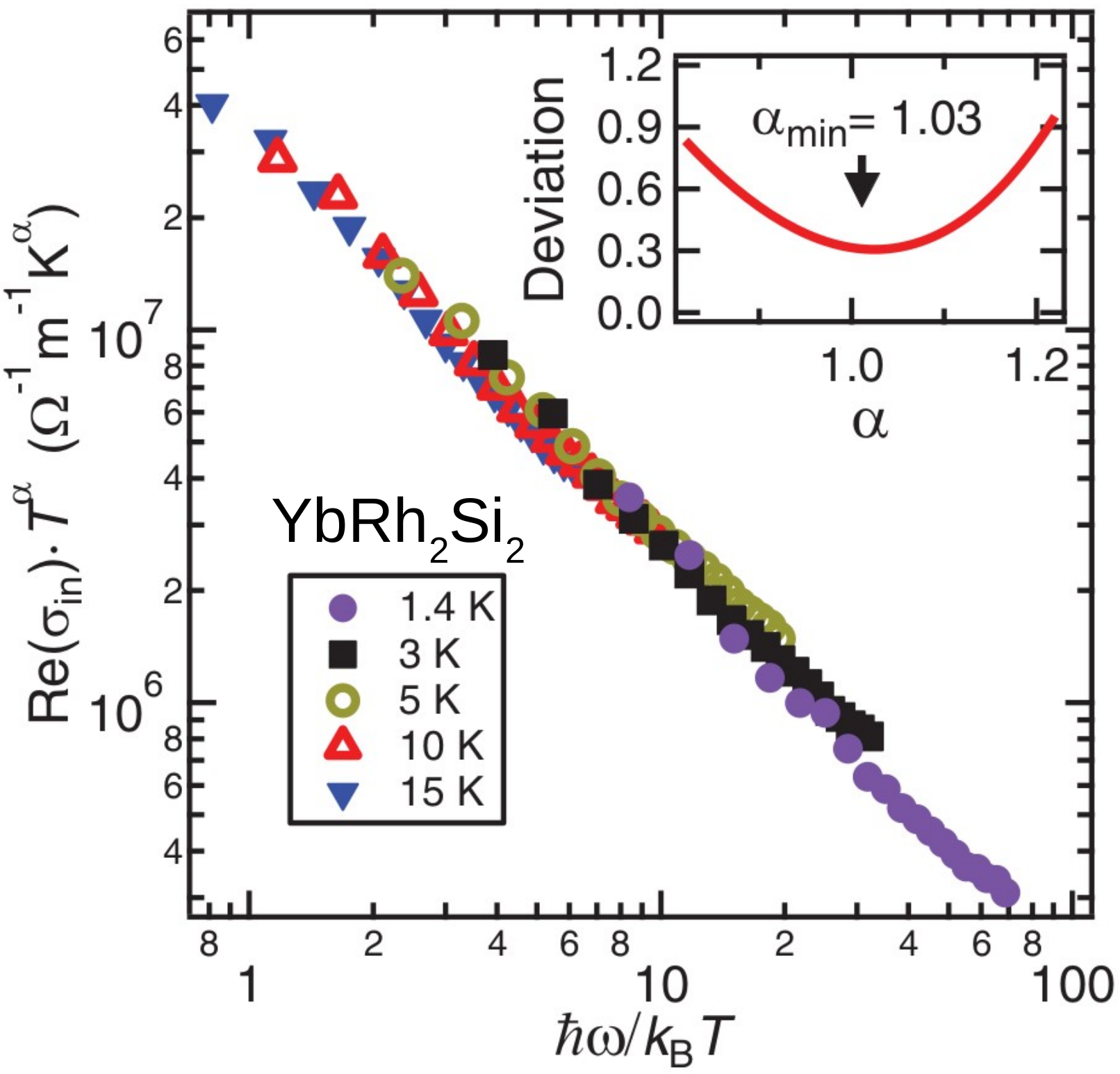}
\vspace{0.2cm}

\hspace{0.2cm}\includegraphics[height=5.5cm]{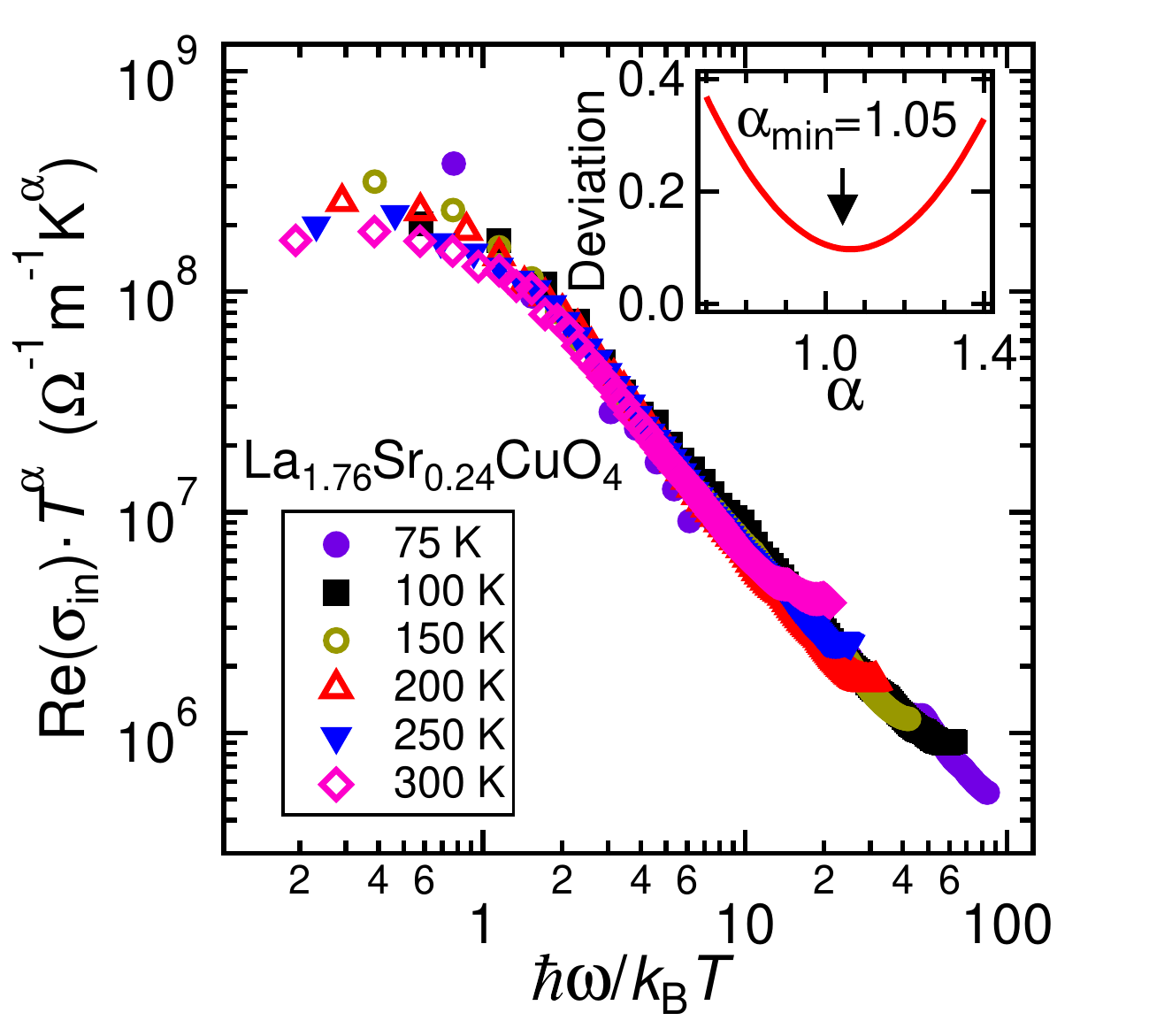}\hspace{-0.2cm}\includegraphics[height=5.5cm]{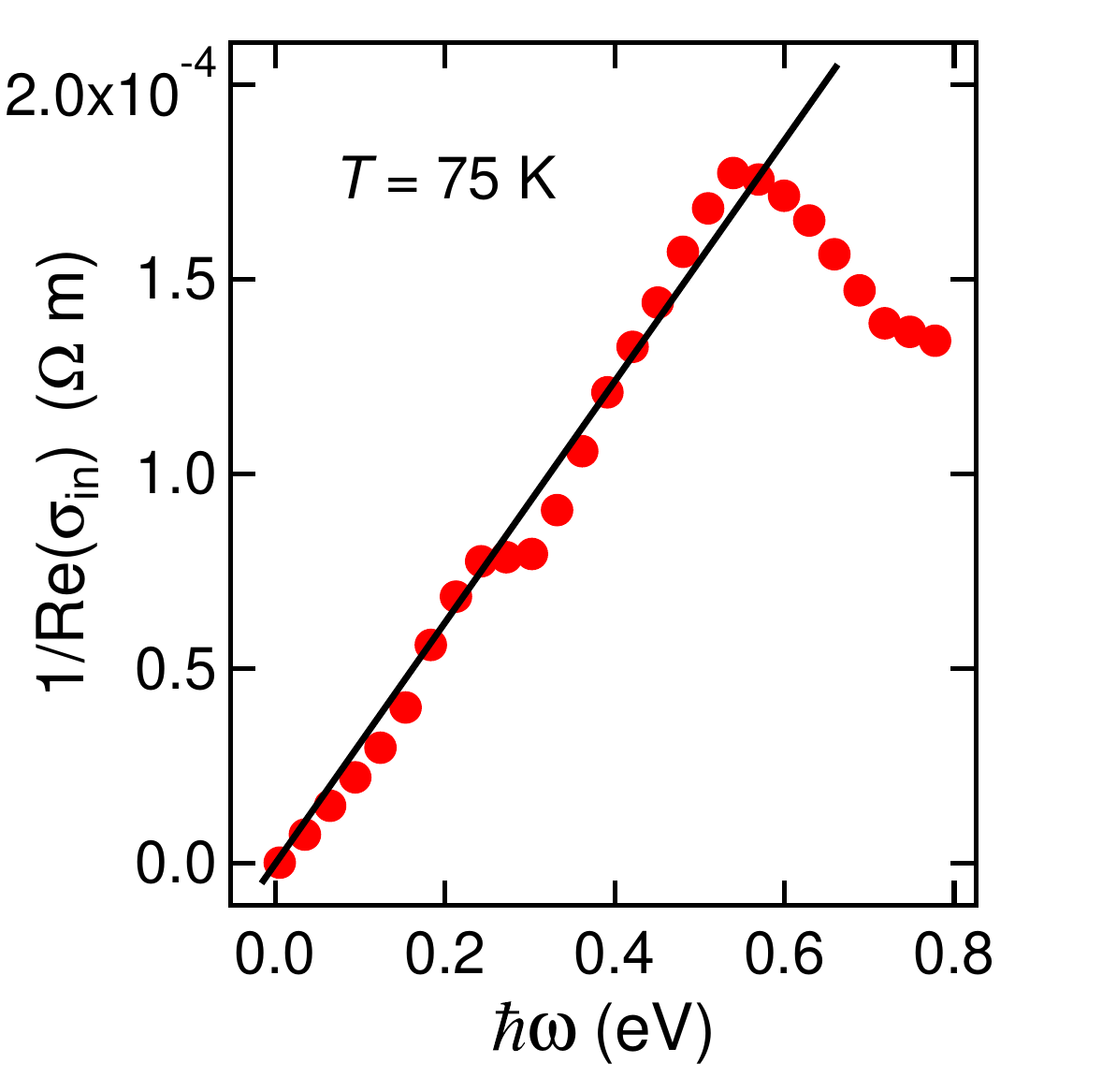}\hspace{-0.2cm}\includegraphics[height=5.5cm]{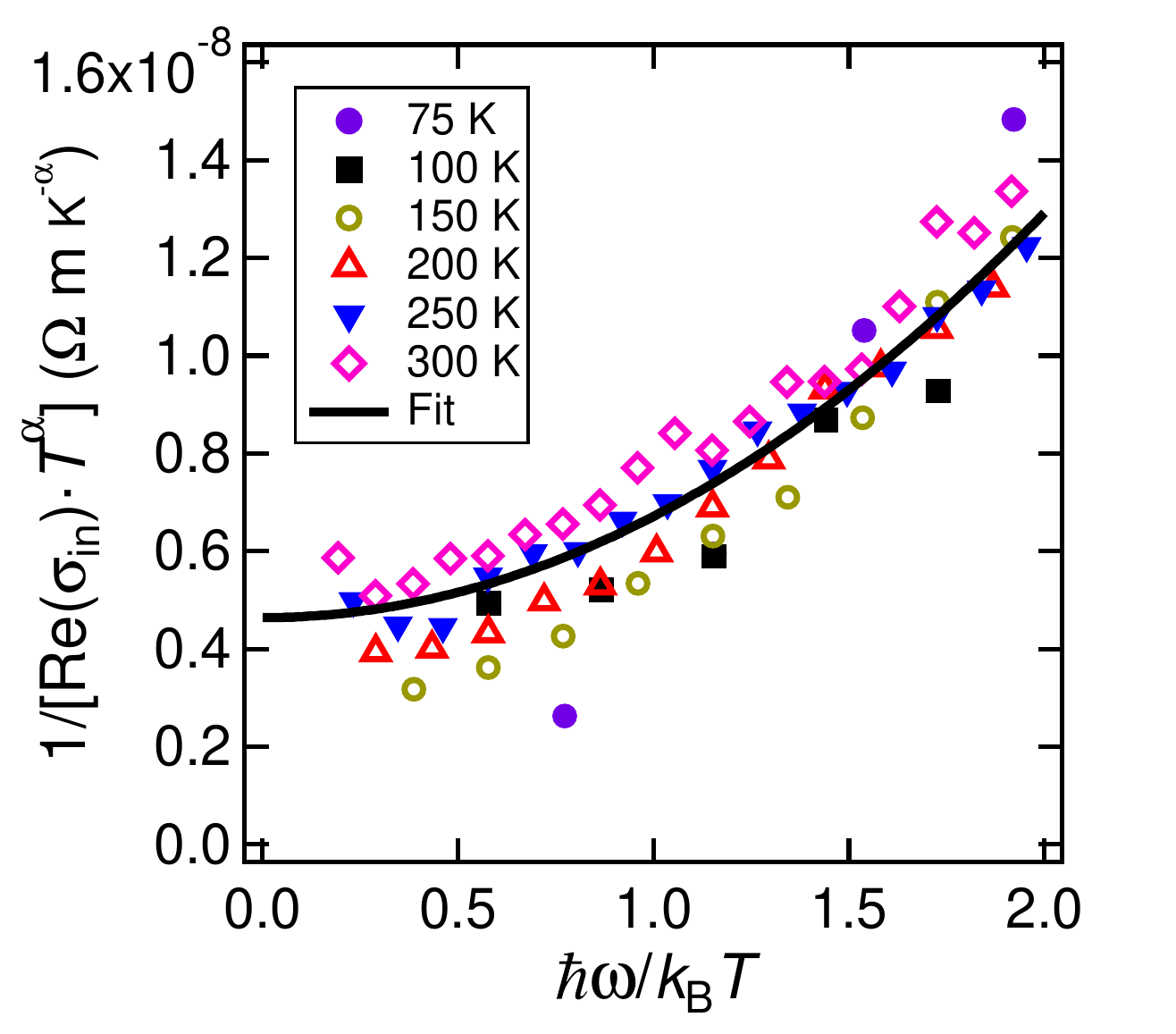}
\end{center}
\vspace{-13cm}

\hspace{0.4cm}{\large\bf{\fontfamily{phv}\selectfont A}}\hspace{9.4cm}{\large\bf{\fontfamily{phv}\selectfont B}}
\vspace{6.6cm}

\hspace{0.25cm}{\large\bf{\fontfamily{phv}\selectfont C}}\hspace{5.5cm}{\large\bf{\fontfamily{phv}\selectfont D}}\hspace{5.2cm}{\large\bf{\fontfamily{phv}\selectfont E}}
\vspace{5.1cm}

\caption{Energy-over-temperature scaling of the real part of the optical
conductivity of \textbf{(A)} the cuprate
Bi$_2$Sr$_2$Ca$_{0.92}$Y$_{0.08}$Cu$_2$O$_{8+\delta}$ and \textbf{(B)} an MBE
film of YbRh$_2$Si$_2$. For the latter, the residual conductivity was subtracted
as explained in \citep{Pro20.1}. Note the different temperature and frequency
ranges, above 100\,K (above the compound's superconducting transition
temperature $T_{\rm c} = 96\,{\rm K}$) and 1.5\,THz (far infrared) in the
former, and above 1.4\,K and 0.23\,THz (THz range) in the latter. In
\textbf{(C)} we replot optical conductivity data of La$_{2-x}$Sr$_x$CuO$_4$ with
$x = 0.24$ from \citep{Mic22.1x}, after subtracting the residual conductivity
(see text), in the same kind of scaling plot as shown in panel B. Using data
below 0.6\,eV, the best scaling collapse is obtained for $\alpha \approx 1$
(inset), just as for YbRh$_2$Si$_2$. \textbf{(D)} Optical resistivity $1/{\rm
Re}[\sigma_{\rm in}(\omega)]$ at 75\,K, which is approximately linear in
frequency up to $0.6\,{\rm eV}$. \textbf{(E)} Fit of the scaled data from panel
C with the scaling function defined in equations\,\ref{eq:omega-T-scaling-via-y}
and \ref{eq:scaling-function-vdM}, as used in \citep{Mar03.1} and
\citep{Mic22.1x}. Panel A is adapted from \citep{Mar03.1}, panel B from
\citep{Pro20.1}. Data in panel C-E are taken from
\citep{Mic22.1x}.}\label{fig:scaling}
\end{figure}

When the energy equivalents of temperature ($k_{\rm B}T$) and frequency
($\hbar\omega$) are similar in magnitude, neither the temperature nor the
frequency dependence alone can characterize the conductivity of a material.
Here, testing scaling relationships in terms of frequency and temperature
becomes of crucial importance. For the cuprate
Bi$_2$Sr$_2$Ca$_{0.92}$Y$_{0.08}$Cu$_2$O$_{8+\delta}$ \citep{Mar03.1}, the
scaling behavior ${\rm Re}(\sigma)\cdot \sqrt{\omega/\omega_0} =
f(\hbar\omega/k_{\rm B}T)$ was demonstrated for $3 k_{\rm B}T < \hbar\omega < 30
k_{\rm B}T$ (not shown here) and a different one, ${\rm Re}(\sigma)\cdot T =
f(\hbar\omega/k_{\rm B}T)$, was observed in the relatively narrow range $0.7
k_{\rm B}T < \hbar\omega < 1.7 k_{\rm B}T$ (figure\,\ref{fig:scaling}A). In
YbRh$_2$Si$_2$, the latter scaling relationship, applied to ${\rm
Re}(\sigma_{\rm in})$, holds in the entire accessed part of the material's
strange metal regime (figure\,\ref{fig:scaling}B). The linear temperature
exponent is consistent with the linear-in-temperature strange metal behavior of
the dc resistivity (figure\,\ref{fig:YRS_1}A) and with the linear-in-frequency
intrinsic ``optical resistivity'' 1/Re$(\sigma_{\rm in})$ at 1.4\,K
(figure\,\ref{fig:YRS_1}B), and indicates its dynamical nature. As further
discussed below, it will be important to see whether the same scaling law holds
also for $\hbar\omega \ll k_{\rm B}T$ (and within the strange metal regime), and
whether an analytical function that describes the data in this regime can be
found.

Note that the scaling analysis of \citep{Mar03.1} was performed with the total
optical conductivity. To examine the influence of the residual resistivity, we
used optical conductivity data of the cuprate La$_{2-x}$Sr$_x$CuO$_4$ with $x =
0.24$ (which is close to the pseudogap critical point) \citep{Mic22.1x} and
preformed the correction using equation\,\ref{eq:ac_Matthiessen}. As in
\citep{Pro20.1}, the residual resistivity $1/\sigma_{\rm res}(\omega)$ was
calculated by assuming that $\sigma_{\rm res}(\omega)$ takes the Drude form,
with the dc value $\rho_{\rm res}=12.2\,\mu\Omega$cm as given in
\citep{Mic22.1x}. The residual scattering rate that appears in $\sigma_{\rm
res}(\omega)$ was calculated using $\tau_{\rm res}=m/(ne^2\rho_{\rm res})$,
where the ratio $m/n$ was estimated from the plasma frequency of around 1\,eV
given in \citep{Uch91.1}, which is similar to that of YBa$_2$Cu$_3$O$_{6+x}$
\citep{Ore90.1}. For ${\rm Re}[\sigma_{\rm in}(\omega)]$ determined in this way
we attempted the same scaling procedure as in \citep{Pro20.1} and found that the
best scaling is achieved for the exponent $\alpha \approx 1$
(figure\,\ref{fig:scaling}C). Some deviation is observed at $\hbar\omega/k_{\rm
B}T < 1$, likely due to the different slope of the dc electrical resistivity at
temperatures above 150\,K \citep{Mic22.1x}, which becomes most evident at low
frequencies. Nevertheless, the scaling holds over an appreciable temperature and
frequency range, which is quite remarkable. Furthermore, at the lowest
temperature of 75\,K, this scaling is anchored by an optical resistivity $1/{\rm
Re}[\sigma_{\rm in}(\omega)]$ that is linear in frequency over the sizable
frequency range of $0 - 0.6\,{\rm eV}$ (figure\,\ref{fig:scaling}D), again in
close similarity with YbRh$_2$Si$_2$ (figure\,\ref{fig:YRS_1}B).

\section{Relaxation rate defined via the scaling form of the optical conductivity}\label{RelaxationScaling}

Finally, we turn to the question of how to extract a relaxation rate from the
scaling function of the real part of the (intrinsic) optical conductivity. On
general grounds, it can be defined in the low-frequency regime $\hbar\omega \ll
k_{\rm B}T$ as
\begin{eqnarray} 
\Gamma = \left[-\frac{\partial\ln{\rm Re}[\sigma_{\rm in}(\omega, T)]}{\partial\omega^2}\right]^{-1/2}_{\omega\rightarrow 0} \, .
\label{eq:Gamma-def}
\end{eqnarray} 
We will show that, provided the strange metal scaling relationship
\begin{eqnarray}
{\rm Re}[\sigma_{\rm in}(\omega,T)] \cdot T^{\alpha}
= f\left ( \frac{\hbar \omega}{k_{\rm B} T} \right ) \quad \mbox{with} \quad \alpha = 1
\label{eq:omega-T-scaling}
\end{eqnarray}
extends to $\hbar \omega/k_{\rm B}T \ll 1$ and such data are available, it may
be used to extract a meaningful $\omega\rightarrow 0$ (but finite $T$) strange
metal relaxation rate. So far, such rates have only been estimated from Drude
(or extended- or multi-Drude) fits, which is inappropriate in the strange metal
regime. By rewriting this scaling form as
\begin{eqnarray}
{\rm Re}[\sigma_{\rm in}(\omega,T)]
= \frac{1}{T \cdot g(y)} \quad \mbox{with} \quad y = \left(\frac{\hbar \omega}{k_{\rm B}T}\right)^2 \, ,
\label{eq:omega-T-scaling-via-y}
\end{eqnarray}
the relaxation rate becomes
\begin{eqnarray}
\Gamma =  \frac{k_{\rm B}T}{\hbar} \, \left ( \frac{\partial\ln g(y)}{\partial y} \right )^{-1/2}_{y\rightarrow 0} \, .
\label{eq:Gamma-expression}
\end{eqnarray}
We now illustrate the procedure of determining a strange metal relaxation rate using the function 
\begin{eqnarray}
g(y) = C(1+A^2 y)
\label{eq:scaling-function-vdM}
\end{eqnarray}
proposed in \citep{Mar03.1} to describe the low-frequency
Bi$_2$Sr$_2$Ca$_{0.92}$Y$_{0.08}$Cu$_2$O$_{8+\delta}$ data (dashed curve in
figure\,\ref{fig:scaling}A, which gave $A=0.77$). Combined with
equation\,\ref{eq:Gamma-expression} this leads to
\begin{eqnarray}
\Gamma = \frac{1}{A} \frac{k_{\rm B}T}{\hbar} \, .
\label{eq:Gamma-vdM}
\end{eqnarray}
Comparing this relation with equation\,\ref{eq:alpha} one can identify $1/A$
with the coefficient $\alpha$.

We now test this analysis scheme with the ${\rm Re}[\sigma_{\rm in}(\omega,T)]$
data of La$_{1.76}$Sr$_{0.24}$CuO$_4$ from figure\,\ref{fig:scaling}C. The fit
with the scaling function of equation\,\ref{eq:scaling-function-vdM} describes
the data fairly well even up to $\hbar \omega /k_{\rm B}T \approx 2$
(figure\,\ref{fig:scaling}E), with $A=0.67$ (and thus the Planckian coefficient
$\alpha_{\tau} = 1.5$). Note, however, that
equation\,\ref{eq:scaling-function-vdM} is only appropriate in the limit $y \ll
1$ (i.e., $\hbar\omega \ll k_{\rm B}T$, but still in the strange metal regime).
It cannot apply to $y \gg 1$, as otherwise it would imply $1/{\rm
Re}[\sigma_{\rm in}(\omega,T)] \sim \omega^2/T$, in contradiction with the
linear-in-$\omega$ behavior seen in figure\,\ref{fig:scaling}D. Indeed, the
lowest-temperature data in figure\,\ref{fig:scaling}E ($75\,{\rm K}$, violet
circles) are the least well described by the fit. It is clear that
higher-resolution data at sufficiently low temperatures, deep in the strange
metal regime, and simultaneously in the regime $\hbar\omega \ll k_{\rm B}T$ are
needed to confirm the scaling function and the value of $A$. Such data will be
highly valuable for any strange metal, to conclusively answer the questions
whether strange metals are ``Planckian''. For the case of YbRh$_2$Si$_2$, this
will require data in the microwave regime.

\section{Relation of temperature and frequency in the Drude model}\label{Drude_UPA}

\begin{figure}[t!]
\begin{center}
\includegraphics[height=5cm]{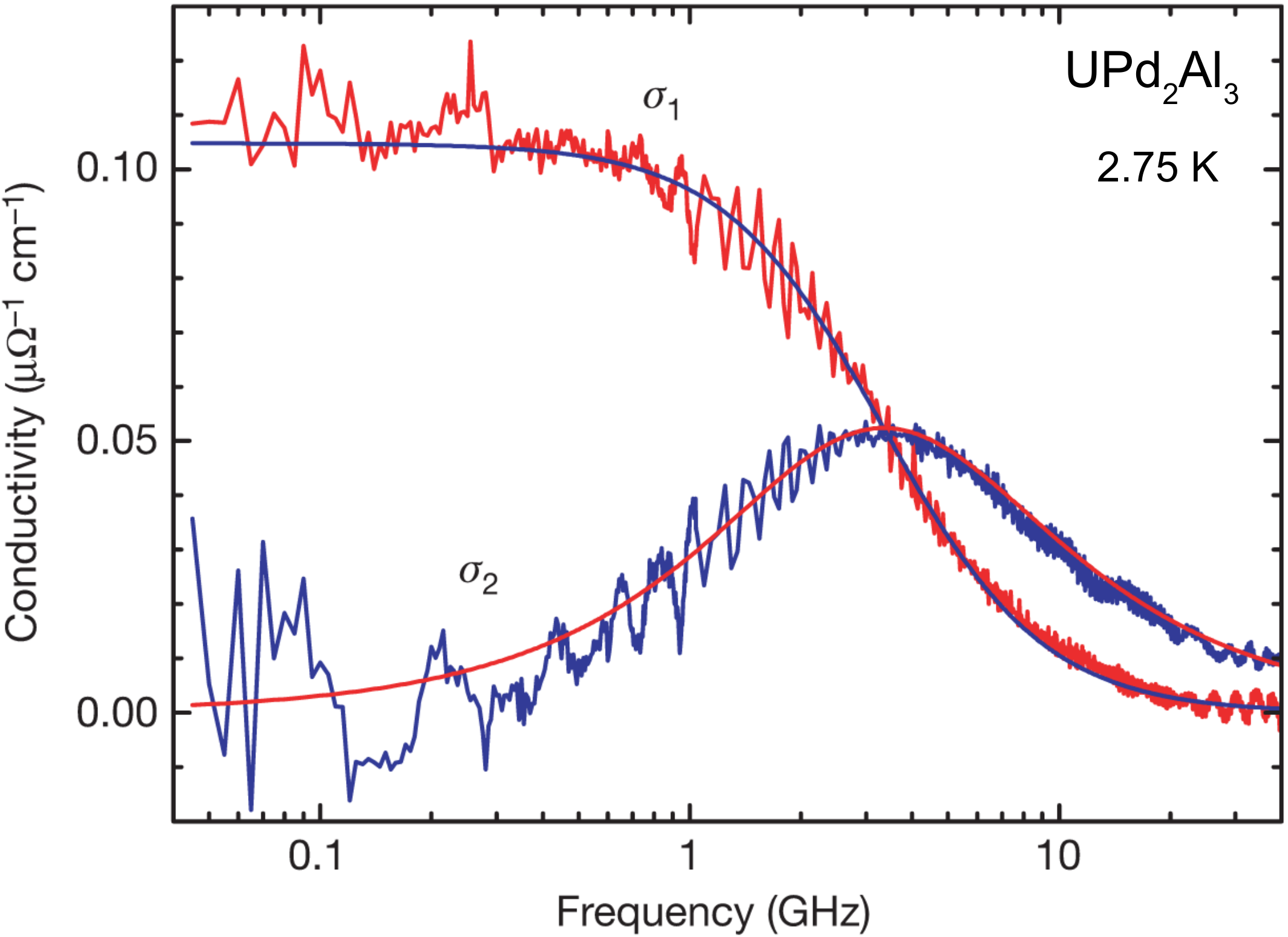}\hspace{0.6cm}\includegraphics[height=5.2cm]{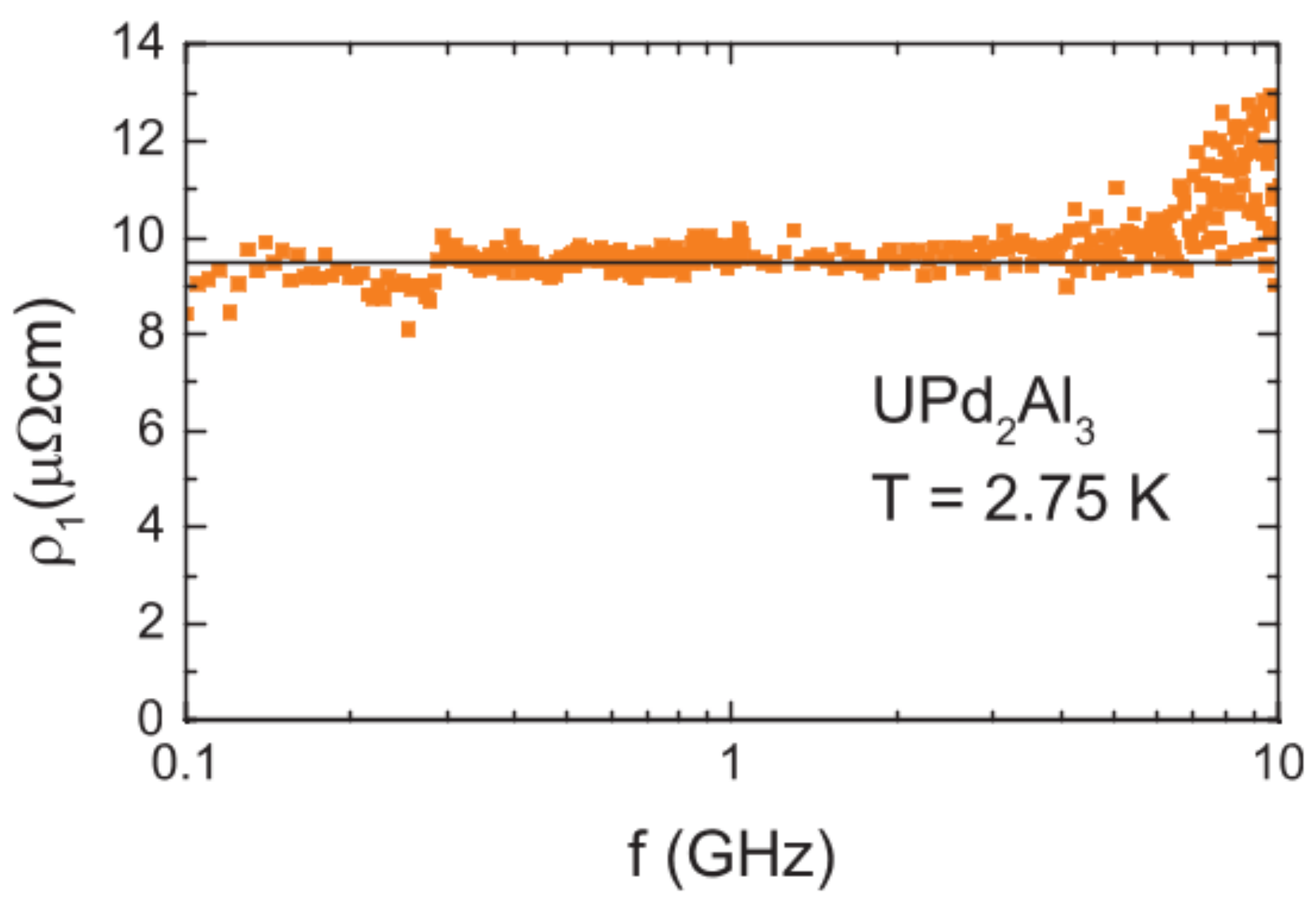}
\end{center}
\vspace{-5.3cm}

\hspace{1.3cm}{\large\bf{\fontfamily{phv}\selectfont A}}\hspace{7.2cm}{\large\bf{\fontfamily{phv}\selectfont B}}
\vspace{4.8cm}

\caption{Optical conductivity of an MBE film of UPd$_2$Al$_3$ at 2.75\,K in the
microwave regime. \textbf{(A)} The real and imaginary parts of the conductivity,
$\sigma_1$ and $\sigma_2$, are shown together with a Drude fit to both
components that yielded the parameters $\sigma_0 = ne^2\tau/m =
0.105\,\mu\Omega^{-1}{\rm cm}^{-1}$ and $\tau = 4.8 \cdot 10^{-11}\,{\rm
s/rad}$. \textbf{(B)} Real part of the resistivity,
$\rho_1(\omega)=\sigma_1(\omega)/|\sigma(\omega)|^2$, of the same data. The
solid line includes the $\omega^2$ term estimated as explained in the text. The
$\omega^2$ upturn is not resolved on the plotted scale. Panels A and B are
adapted from \citep{Sch05.2} and \citep{Sch13.3},
respectively.}\label{fig:UPd2Al3}
\end{figure}

As a side aspect, we comment here on the relation between temperature and
frequency in Fermi liquids. That a temperature dependent scattering rate should
go along with a corresponding frequency dependence was realized early on, and
formulated for the Fermi liquid case as \citep{Gur59.1}
\begin{equation}
\frac{1}{\tau} = \frac{1}{\tau_{\rm res}} + a (k_{\rm B}T)^2 + b (\hbar \omega)^2 \quad \mbox{with} \quad \frac{a}{b} = 4 \pi^2  . \label{eq:Gurzhi}
\end{equation}
Surprisingly, according to \citep{Sch13.3}, this relation has remained untested.
The challenge is to resolve the frequency dependence of equation \ref{eq:Gurzhi}
in the Fermi liquid regime, i.e., at sufficiently low temperatures and
frequencies. As shown in \citep{Sch13.3}, this has not even been possible in
heavy fermion compounds, where the prefactors are strongly enhanced
\citep{Kad86.1,Jac09.1}. In a tour de force effort, the optical conductivity of
MBE grown thin films of the heavy fermion compound UPd$_2$Al$_3$ was measured at
low temperatures in the microwave regime \citep{Sch05.2}. The data at 2.75\,K
(figure \ref{fig:UPd2Al3}A), however, closely follow a simple Drude law, with a
frequency independent scattering rate. As a consistency check, the magnitude of
the third term in equation \ref{eq:Gurzhi} was estimated from the $A$
coefficient of a bulk polycrystalline sample [$A = 0.23\,\mu\Omega{\rm cm}/{\rm
K}^2$  \citep{Dal92.1} which, assuming the validity of the Kadowaki--Woods ratio
\citep{Kad86.1}, is consistent with the Sommerfeld coefficient $\gamma_0 =
150\,\rm{mJ/mol K}^2$ \citep{Gei91.1}]. The relative increase of the inverse
optical conductivity (figure \ref{fig:UPd2Al3}B) due to the third term was found
to be well below the resolution limit of the experiment [less than 0.1\% at
20\,GHz \citep{Sch21.1}]. Using the somewhat larger value $A =
0.51\,\mu\Omega{\rm cm}/{\rm K}^2$ of a UPd$_2$Al$_3$ MBE film \citep{Hut93.1}
would not have changed the conclusion.

Another heavy fermion compound on which optical conductivity measurements have
been performed to rather low frequencies (10\,GHz), and down to 1.2\,K, is
CeAl$_3$ \citep{Awa93.1}. Fermi liquid behavior in the dc resistivity, with a
very large $A$ coefficient of $35\,\mu\Omega{\rm cm/K}^2$, was reported below
300\,mK \citep{And75.3}. The scattering rate (and the effective mass), extracted
using an extended Drude model (see section\,\ref{discussion}), indeed show
pronounced frequency dependence at low temperatures \citep{Awa93.1}. Our
analysis of these data reveals that, even at the lowest frequencies and
temperatures, no clear $\omega^2$ dependence is seen. Assuming that the two
lowest-frequency data points at 1.2\,K represent the slope of an $\omega^2$
dependence, we estimate $a/b \approx 0.24$, much smaller than $4 \pi^2$. Data at
even lower frequencies and temperatures will be needed to confirm this result.

\section{Summary, discussion, and outlook}\label{discussion}

In this perspective paper, we have examined published optical conductivity data
of various materials to address the question whether strange metal behavior is
captured by Planckian dissipation. The idea of using the optical conductivity
instead of the dc resistivity, as was done in previous work
\citep{Bru13.1,Leg19.1,Cao20.1,Ghi21.1,Gri21.1,Mou21.1,Tau22.1}, was to remove
the uncertainty created by estimating the charge carrier concentration and
effective mass (or, more precisely, their ratio $n/m$). Using the Drude form of
the optical conductivity can, in principle, overcome this problem because here
the scattering time $\tau$ enters not only in a product with $n/m$ (as in the dc
conductivity $\sigma = ne^2\tau/m$) but also in a second term. This allows one
to fit $n/m$ and $\tau$ independently. The ratio $\alpha$ of the experimental
(inelastic) scattering rate $1/\tau_{\rm in}$ and the Planckian scattering rate
$1/\tau_{\rm P} = k_{\rm B}T/\hbar$ can then be determined.

We tested the method with high-temperature data of simple metals and
semiconductors in the range where the resistivity is governed by a
linear-in-temperature scattering rate due to electron-phonon scattering, and
found it to be reliable. Notably, $\alpha$ can be obtained in two different
ways, directly from $\tau_{\rm in}$ (which we then call $\alpha_{\tau}$) or by
combining $n/m$ and the linear-in-temperature dc resistivity coefficient $A'$
(which we call $\alpha_{n/m}$), and both gave very similar results for the
simple metals.

We then attempted to apply the method to the extreme strange metal
YbRh$_2$Si$_2$. However, as already pointed out in \citep{Pro20.1}, the Drude
form of the optical conductivity fails to describe the data at low temperatures
and frequencies, which we would have deemed the most appropriate range to
characterize the compound's strange metal state. To achieve reasonable fits, the
fitting range has to be constrained to sufficiently high temperatures and
frequencies. In this case, a much more modest mass renormalization results than
what is obtained by using the dc resistivity at low temperatures, even at tuning
parameter values far away from the quantum critical value \citep{Tau22.1}. As a
result, both $\alpha_{\tau}$ and $\alpha_{n/m}$ are much larger than what was
found from the dc resistivity analysis. Furthermore, the large discrepancy
between $\alpha_{\tau}$ and $\alpha_{n/m}$ resulting from this procedure points
to the need for alternative approaches---not based on dc or ac Drude models---to
classify strange metal behavior.

Some researchers have used the ``extended'' Drude model \citep{Dre02.2,Mic22.1x}
to describe optical conductivity data of strongly correlated electron systems.
It assumes a complex frequency-dependent relaxation rate $1/\tau(\omega) =
1/\tau_1(\omega) + i/\tau_2(\omega)$ and, via
\begin{equation}
\frac{1}{\tau_1(\omega)} = \frac{1}{\tau^*(\omega)} \cdot \frac{m^*(\omega)}{m}  \quad\quad \mbox{and} \quad\quad  \frac{1}{\tau_2(\omega)} = \omega \cdot [1-\frac{m^*(\omega)}{m}] \; , \label{eq:Drude_ac1}
\end{equation}
a frequency-dependent effective mass related to it. When introduced into
equation\,\ref{eq:Drude_ac}, this can be brought into the generalized Drude form
\begin{equation}
\sigma(\omega) = \frac{n e^2\tau^*(\omega)}{m^*(\omega)} \frac{1}{1-i\omega\tau^*(\omega)} \; .  \label{eq:Drude_ac3}
\end{equation}
Plotted and analysed are typically the quantities
\begin{equation}
\frac{1}{\tau_1(\omega)} = \frac{n e^2}{m} \frac{\sigma_1(\omega)}{|\sigma(\omega)|^2} \label{eq:Drude_ac4}
\end{equation}
and
\begin{equation}
\frac{m^*(\omega)}{m} = \frac{n e^2}{m} \frac{\sigma_2(\omega)/\omega}{|\sigma(\omega)|^2} \; , \label{eq:Drude_ac5}
\end{equation}
but this requires the knowledge of the (unrenormalized) plasma frequency
$\omega_{\rm p}$ (see equation\,\ref{eq:plasma}). At low frequencies and
temperatures, residual scattering may play an important role and should be
subtracted, as described in section\,\ref{Drude}. The charge carrier
concentration $n$ is assumed to be frequency independent in this approach, which
may not be true in heavy fermion strange metals
\citep{Pas04.1,Shi05.1,Fri10.2,Cus12.1,Jia15.1,Mar19.1} and related materials
\citep{Ani02.1,Oik15.1,Bad16.1,Cao18.1,Jia20.1}. More generally, in a strange
metal, any a priori assumption on $n$, $m$, and $\tau$ limits the generality and
may bias the conclusions. Forcing (extended) Drude forms to the data bears the
risk to overlook the essential physics. Strange metals may be governed by exotic
excitations or even the absence of any well-defined quasiparticles
\citep{Si01.1,Col01.1,Sen04.1,Phi11.1,Cha18.1,Pat18.1,Kom19.1,Pep21.1,Cai20.2,Cha20.1,Guo20.1,Bal20.1,Lee21.1,Els21.2,Wan22.1,Cap22.1},
so they defy description by the above models and require alternative approaches.

We consider scaling analyses of the (intrinsic) optical conductivity in the
strange metal regime to be the most promising way forward. To reliably determine
the functional form of the scaling, it will be essential to access, with
high-resolution data, both the regime $\hbar\omega/k_{\rm B}T \gg 1$ and the
regime $\hbar\omega/k_{\rm B}T \ll 1$. From such scaling functions, a
generalized (non-Drude) relaxation rate can then be determined, and compared to
theoretical expectations for strange metals. In strange metals with low energy
scales, such as the heavy fermion compounds, the latter will require challenging
low-temperature experiments in the microwave regime. When high-quality thin
films are available, this can in principle be achieved with the broadband
Corbino technique \citep{Sch05.1}, as demonstrated for the Fermi liquid heavy
fermion compound UPd$_2$Al$_3$ \citep{Sch05.2} (figure \ref{fig:UPd2Al3}). We
note that even in state-of-the-art THz experiments on MBE grown thin-films of
YbRh$_2$Si$_2$ \citep{Pro20.1}, this regime was not accessed. This calls for
future studies to advance the field.

\section*{Conflict of Interest Statement}

The authors declare that the research was conducted in the absence of any
commercial or financial relationships that could be construed as a potential
conflict of interest.

\section*{Author Contributions}

XL performed the Drude and scaling analyses, SP conceived the work and wrote the
paper, with input from XL, JK, and QS. All authors contributed to the
discussion.

\section*{Funding}

XL acknowledges support from the Caltech Postdoctoral Prize Fellowship and the
IQIM. JK acknowledges support from the Robert A.\ Welch Foundation through Grant
No.\ C-1509. QS acknowledges support from the Air Force Office of Scientific
Research under Grant No.\ FA9550-21-1-0356 and the Robert A.\ Welch Foundation
under Grant No.\ C-1411. SP acknowledges funding from the European Union's
Horizon 2020 Research and Innovation Programme under Grant Agreement no 824109
and from the Austrian Science Fund (FWF Grants 29296-N27 and I5868-N--FOR 5249 -
QUAST). SP and QS acknowledge the hospitality of the Aspen Center for Physics,
which is supported by NSF grant No.\ PHY-1607611.

\section*{Acknowledgments}

We acknowledge fruitful discussions with Peter Armitage, Antoine Georges,
Patrick Lee, Subir Sachdev, Marc Scheffler, T.\ Senthil, Mathieu Taupin, and
Eric van Heumen. 



\end{document}